# Temperature dependence of the resistance switching effect studied on the metal/YBa$_2$Cu$_3$O$_{6+x}$ planar junctions

Milan Tomasek[*,1], Tomas Plecenik[1], Martin Truchly[1], Jaroslav Noskovic[1], Tomas Roch[1], Miroslav Zahoran[1], Stefan Chromik[2], Mariana Spankova[2], Peter Kus[1] and Andrej Plecenik[1]

1 - Faculty of Mathematics, Physics and Informatics, Comenius University, Bratislava
2 - Institute of Electrical Engineering, Slovak Academy of Sciences, Bratislava

**Abstract**

Resistive switching (RS) effect observed in capacitor-like metal/insulator/metal junctions belongs to the most promising candidates for next generation of memory cell technology. It is based upon a sudden change of the junction resistance caused by an electric field applied to the metal electrodes. The aim of this work was to study this effect on the structure metal/YBCO$_6$/YBCO$_7$, where YBCO$_7$ is a metallic phase and YBCO$_6$ is an insulator phase which arises spontaneously by out-diffusion of oxygen from a few nanometers wide YBCO surface layer. Oriented YBa$_2$Cu$_3$O$_7$ thin films were prepared by the method of magnetron sputtering and consequently planar structures with metal-YBCO junction were made by the means of the optical lithography, ion etching and vacuum evaporation. On these junctions we have studied the temperature dependence of the RS effect with I-V and dI/dV-V transport measurements down to liquid He temperature. We have determined temperature dependence of the RS effect threshold voltage in the temperature range 100-300 K and showed that this dependency is compatible with common idea of oxygen ions migration under electric field within the YBCO surface layer.

[*] corresponding author e-mail: milan.tomasek@gmail.com

**Introduction**

It is estimated that the conventional silicon-based technology will reach its limit within one decade as its scale is still shrinking and new effects arise in the nanometer scale. Effects like electromigration or dielectric breakdown are unavoidable in such scale and markedly trim the lifetime of electronic elements. To avoid drawbacks of these effects, future technologies should use them for their own benefit. One of the most promising candidates for next generation of memory cell technology based on such effects is so called Resistive RAM (ReRAM) memories[1-3]. They operate on the resistive switching (RS) phenomenon which is typically observed in capacitor-like metal/insulator/metal geometry and it is based upon an abrupt change of the overall resistance of such structure caused by an electrical field applied across the insulating part. Electric field causes reversible structural changes in the active dielectric part what leads to switching between "low-resistive state" (LRS) and "high-resistive state" (HRS). The switching effect is observed in large variety of materials but it is studied mainly on transition-metal oxides and especially in simple two-component oxides, such as $TiO_2$, NiO or $Al_2O_3$ which typically exhibits the unipolar RS effect[4-6]. The switching phenomenon remains less studied on complex oxide systems such as high-$T_c$ superconductors with perovskite structure (YBCO[7], BSCCO[8], LCMO[9], etc.) although these materials are very convenient for study of the switching mechanism since their physical properties (resistivity vs. oxygen vacancy concentration, oxygen diffusivity, etc.) are very well known thanks to extensive prior research of high-$T_c$ superconductivity[10,11]. Compared to unipolar RS effect observable on the symmetrical structures metal/insulator/metal with width of the insulating part a few tens of nanometers, the RS effect on the perovskite materials is observable on the asymmetrical structures metal-1/insulator/metal-2 where a few nanometers wide insulating part is a result of artificially (e.g. different doping) or naturally (e.g. Schottky barrier) created layer at the interface of two different metals or metal and semiconductor. Such structures exhibit bipolar RS effect[12,13].

We have already reported our measurements of the RS effect on the Ag/YBCO and PtIr/YBCO interfaces elsewhere[14] where we have compared this effect in the micro and the submicro scale of the junction contact area. It was observed that RS effect is occurring on the metal/YBCO interface. It is well known that on the surface of the YBCO few nanometers wide degraded layer with semiconducting properties arises within several hours at standard conditions[10,15]. It is a layer with high concentration of oxygen vacancies present due to spontaneous out-diffusion of the oxygen from the material. Insulating character of the degraded layer is direct consequence of the relation between the resistivity and the oxygen content in the unit cell which is that the resistivity is exponentially increasing with decreasing oxygen content[16]. It is believed that the RS effect is occurring in this degraded layer as an effect of the oxygen atoms rearrangement near the interface. This concept is very common among the authors[7,8,17,18]. Particularly, in the YBCO this hypothesis is very well supported by measurements of the density of states of the YBCO electrode in the superconductive state[18]. Transition from the metal-insulator-superconductor tunneling regime in the HRS to regime with appearance of the Andreev reflection typical for metal-superconductor interface in the LRS could be interpreted only as the consequence of the oxygen drift towards the metal electrode. Thus, the oxygen deficient insulating YBCO phase could come to oxygen enriched superconductive phase.

The main point of interest in the physics involved behind the switching mechanism today is the origin of the oxygen movement. There are two apparent concepts on this issue. One is that the oxygen movement is a consequence of the electrically stimulated Mott transition[19] in the dielectric phase on the interface (due to trapping or releasing the charges from the barrier) which in combination with strong correlation effects leads to rearrangement of the oxygen[8,20-22]. In our case, dielectric phase of oxygen-depleted $YBa_2Cu_3O_6$ is known as the antiferromagnetic insulator where free movement of charge carriers is forbidden due to strong spin localization in ordered antiferromagnetic spin lattice. Adding oxygen into unit cell, the valence state of Cu metal cations is changed and spin lattice is disturbed what results in release of charge carriers and metal character of the $YBa_2Cu_3O_7$[11]. Thus the conductivity and spin ordering are strongly linked with the oxygen content in the unit cell what means that electrically stimulated Mott transition at some threshold level could lead to rearrangement of the spins and thus the oxygen atoms. After this happens, the new oxygen structure together with resistivity state is preserved at voltages below the Mott transition threshold voltage. Second concept is much more straightforward and says that the oxygen rearrangement is a direct consequence of the oxygen ions electro-migration under electric field. Thus, at the positive (negative) polarity on the metal (YBCO) oxygen anions are moving toward the metal electrode while filling the oxygen vacancies in the insulating layer (insulator→metal) and at the opposite polarity they are moving toward the YBCO electrode leaving the oxygen vacancies at the interface (metal→insulator)[7,17]. Difference between these concepts is in the roles of the cause and the effect. In the first one, the electronically stimulated metal-insulator (Mott) transition is conserved by the rearrangement of the oxygens. In the second one, the electromigration of the oxygen anions leads to metal-insulator transition due to different oxygen doping in the interface layer. Anyway in both cases, the mobility of the oxygen is based on the thermal diffusivity of the oxygen in zero electric field which strongly depends on the temperature. Therefore the temperature changes should have some effect also on the RS effect parameters. Recently, Acha studied the RS phenomenon in contacts formed by the ceramic YBCO sample and metal counter-electrode at various temperatures in the range 300 K – 440 K[7]. In this paper, we present our measurements of the RS effect on the metal/YBCO interfaces in the temperature range 100 K – 300 K. In both cases, with increasing temperature the threshold switching voltage was decreasing. We discuss this observation in the context of the electro-diffusion transport mechanism of the oxygen ions.

**Experimental**

Single-crystalline $LaAlO_3$ (001) substrates offering small lattice and thermal mismatches were used to obtain c-axis oriented $YBa_2Cu_3O_7$ films with thickness 200 nm. Detailed description of the deposition process by magnetron sputtering was described elsewhere[14]. The resistance versus temperature characteristics of these YBCO films exhibited metallic dependence above critical temperature ($T_c$ ~ 93 K) with transition downset at about 87 K. The 100 μm wide strip electrodes were formed by the optical lithography and subsequent Ar ion etching from the YBCO thin film. The next step consisted in $SiO_2$ layer deposition by the means of the vacuum evaporation technique and removal of the photoresist. Then, the lift-off photolithography was used to obtain 5, 10, and 15 μm wide Ag strips across the planarized YBCO electrode. The final planar structure is shown schematically in Fig. 1.

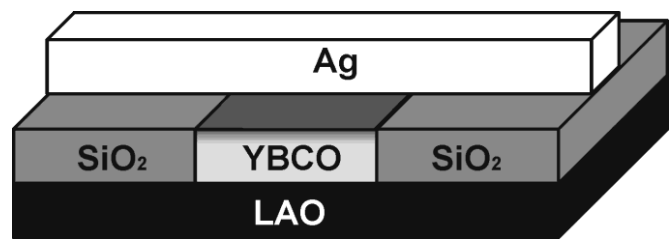

*Fig. 1 – Schematic representation of the Ag/YBCO planar junction. The 100 μm wide silver upper-electrode lies across the 10 μm wide YBCO bottom-electrode which is planarized by the SiO₂ insulating layer.*

I-V measurements of the prepared junctions were performed in the four-contact configuration according to Fig.2 using the Keithley 6430 SourceMeter in the current-biased mode with bias in a form of the saw sweep with the period ranging from 10 to 1000 seconds. The differential dV/dI-V characteristics were measured by the low-frequency phase-sensitive detection technique in a bridge configuration and the dI/dV-V characteristics were obtained by numerical inversion of dV/dI-V. In all measurements, the YBCO electrode was biased and the Ag counter-electrode was grounded. The temperature dependence was studied in the range from 4.2 K to 300 K in a transport He Dewar container.

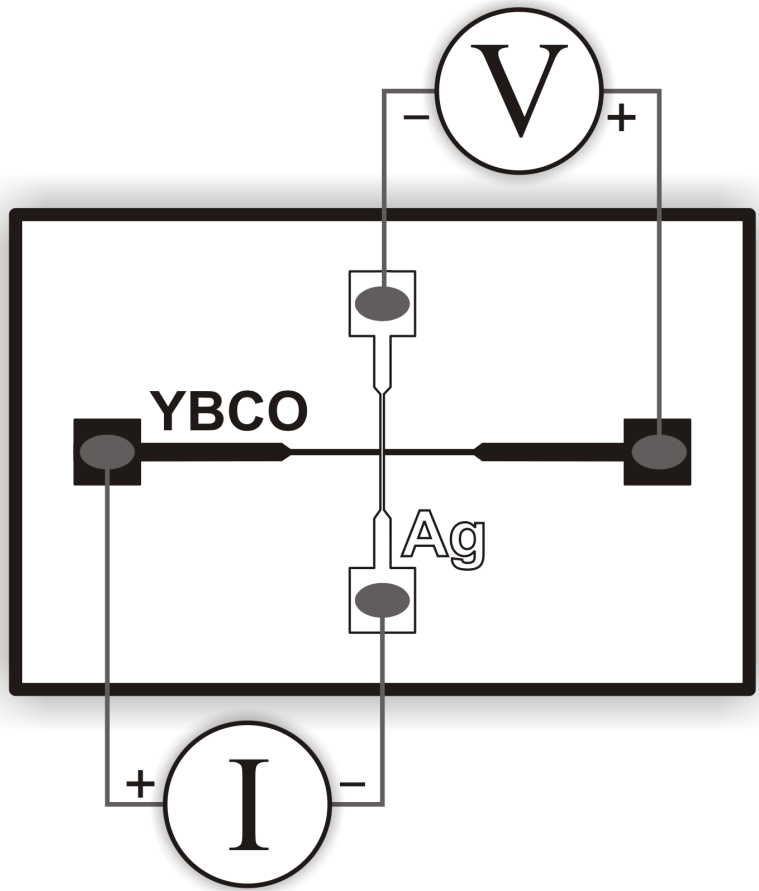

*Fig. 2 – Transport measurement of the Ag/YBCO planar junction. Current flowing across the junction was controlled in a form of the saw sweep with period up to 1000 seconds while the voltage drop on the junction was measured.*

**Results and discussion**

Typical I-V characteristics measured on the planar Ag/YBCO junction is shown in the Fig.3. It demonstrates switching of the junction between the high-resistive state (HRS) and the low-resistive state (LRS) which happened above certain threshold voltage. The resistance switching was reversible and each resistive state was stable for at least hours. HRS to LRS switching voltages was confined to negative polarities and backward switching voltages to positive polarities on the YBCO electrode. The switching process was much more stable and the resistive states were reproducible after introducing the planarization step with the $SiO_2$ layer into preparation of the planar junctions. We have done measurements down to 4.2 K but the switching effect was observed only in the temperature range 100 K – 300 K as below 100 K the planar junctions were destroyed by high current densities.

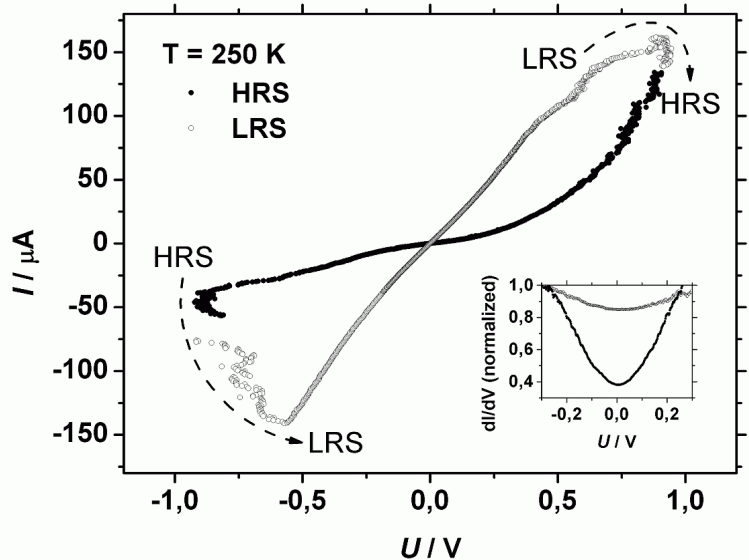

*Fig.3 – Typical I-V characteristics of the Ag/YBCO planar junction. I-V characteristics measured in the current bias mode with two resistive states at the temperature 250 K. Changes in resistive state occurred above certain threshold level. Resistance measured at 1 µA is ~50 kΩ for the high-resistive state (HRS) and ~5 kΩ for the low-resistive state (LRS). Inset: typical differential (dI/dV-V) characteristics for HRS and LRS measured in narrow voltage range well below threshold voltage level.*

The initial state of the junctions was a high-resistive one what could be explained by a well known fact that the surface layer of the YBCO degrades. Insulating character of this layer is given by a high concentration of the oxygen vacancies. If this insulating layer is thin enough (few nanometers) it operates as a tunneling barrier for electrons so the Ag/YBCO junction transport properties can be studied in the frame of the electron tunneling theory. The inset of the Fig.3 shows typical differential characteristics of the HRS and LRS measured at voltages well below the threshold switching voltage. They exhibit parabolic behavior typical for elastic tunneling across the structure metal-insulator-metal. Fitting these characteristics with elastic tunneling formula[23] we have found that the typical width of the insulating barrier at Ag/YBCO interface for the HRS is 2.3 nm and for the LRS it changes down to 1.3 nm. The dominant influence of the tunneling barrier on the transport properties is strongly supported by the correlation between resistance $R$ of the junction and the fitted barrier width $d$ which is shown in the Fig.4. In the tunneling regime, at small biases ($U \ll \phi/e$) the tunneling current is proportional to $I \sim U\exp(-d\sqrt{\phi})$ where $\phi$ is the barrier height. This demands exponential dependence of the tunneling junction resistance on the barrier width. Scatter in this dependence could be interpreted as effect of different barrier heights $\phi$ in different resistance states of the contacts which varied in the range from 0,6 eV to 1,0 eV at the temperature 300 K. This reasoning confirms the existence of the few nanometers wide tunneling barrier at Ag/YBCO interface.

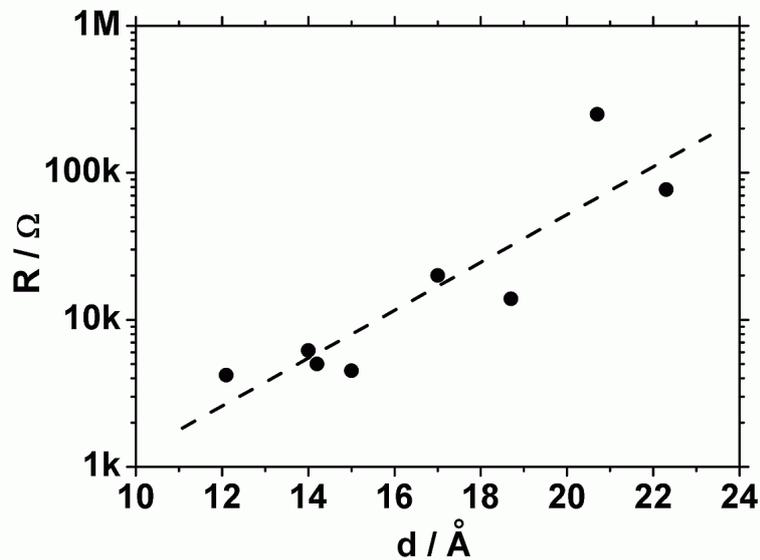

*Fig.4 – Tunneling barrier at the Ag/YBCO interface. Resistance R of the Ag/YBCO junction at the temperature 300 K as the function of the fitted barrier width d in the semi-log plot. Classical tunneling barrier with given width d and constant height is characterized by the exponential relation[23] $R \sim exp(d)$.*

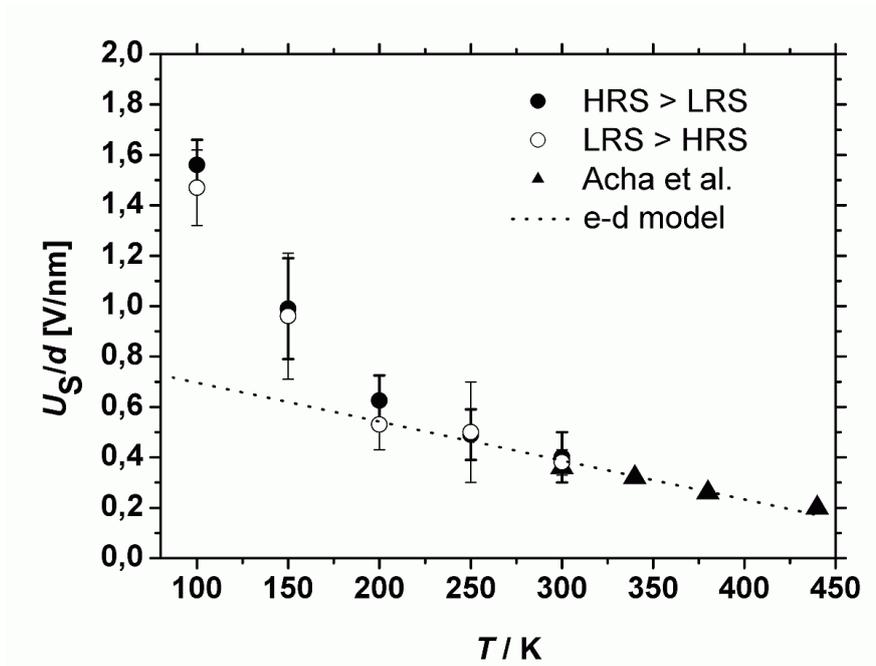

*Fig.5 – Temperature dependence of the switching threshold voltage. Experimantal data above 300 K were taken from Acha et al.[7]. While Acha et al. measured linear relation between the switching voltage and the temperature in the range 300 – 440 K, our dependence in the range 100 – 300 K is obviously non-linear. Linear part of the dependence in the range 200 – 440 K is fitted with the electro-diffusion (e-d) model presented in the text.*

Main feature of the RS effect is its threshold character and changes occurring above certain threshold voltage. We have combined our measurements on the Ag/YBCO junctions together with measurements of Acha et al. on their Au/YBCO junctions[7]. Joining all these data we have received the temperature dependence of the threshold switching voltage $U_S$ in the temperature range 100 – 440 K shown in the Fig.5. While Acha et al. measured linear relation between the switching voltage and the temperature (300 – 440 K), our dependence is obviously non-linear (100 – 300 K). In our measurements, the absolute values of the threshold voltages for switching from the HRS to the LRS and for backward switching from the LRS to the HRS were quantitatively different at given temperature but the temperature dependences were qualitatively the same. In fact, the threshold switching voltages differ only in a constant multiplicative factor. Normalizing the switching voltages with the typical barrier width for the HRS (2.3 nm) and for the LRS (1.3 nm) the temperature dependences have shown to be quantitatively the same. From this we can conclude that the switching mechanism seems to be similar at both polarities as it happens at constant electric field intensity for given temperature (typically $5.10^6$ V/cm). To match Acha's temperature dependence to ours at common point (300 K), it was necessary to normalize their data by a larger factor from what it follows that their degraded layer was much thicker ($d \sim 25$ nm). In fact, the width of the degraded layer is strongly depended on the maximum temperature involved in the process of thin films creation. The degraded layer is thicker as the higher temperature is involved and it could be up to hundreds of nanometers[24]. In our case, before deposition of the upper metal electrode we have performed the ion etching of the YBCO surface what lead to complete removal of the degraded layer and its new creation at room temperature when oxygen diffusivity is much lower then at typical temperatures during the YBCO thin film annealing process (500 °C). At room temperatures the degraded layer of just a few nanometers appears[25].

We suppose that a key role in the RS effect plays the oxygen rearrangement just in the degraded layer on the interface. It seems reasonable as the resistivity of the oxygen deficient YBCO ($\rho > 1000$ mOhm.cm) is a few orders higher than that of the metallic phase ($\rho < 1$ mOhm.cm) located behind the degraded layer. Hence almost all voltage bias between metal electrodes drops on the interface and the electric field is concentrated in the thin degraded layer and can reach intensities up to $10^7$ V/cm. In this case, the mobility of the oxygen ions is significantly enhanced by the electrical field and highly overcomes the basal diffusive mobility due to thermal excitations. In this case the movement of the oxygen ions must be studied in the electro-diffusion model. We describe the basal oxygen motion in absence of the electric field in the frame of the barrier diffusion model which is based on one dimensional system of narrow energetic barriers with height $U_a$ (activation energy for oxygen motion) with separation $\lambda$. Then the rate of oxygen jumps over activation barriers between two adjacent oxygen sites 1 and 2 is given as[15]:

(1) $r_{1 \to 2} = r_{2 \to 1} = v = v_0 \exp(-U_a/k_B T)$

where $v_0$ is the attempt frequency. Applying the voltage $U$ on such system we expect that the electric field with the magnitude $E \approx U/d$ is concentrated on the insulating layer with thickness $d$. Electric field causes acceleration of double charged oxygen ions on the distance $\lambda$ between the oxygen sites, gain of the energy $2e\lambda E$ for the $O^{2-}$ ion what leads to enhanced probability for occurrence of the overcomes in the direction of the electric force comparing with opposite one:

(2) $r_{1\to 2} = v_0 \exp\left(-\frac{U_a - 2e\lambda E}{k_B T}\right) > r_{2\to 1} = v_0 \exp\left(-\frac{U_a + 2e\lambda E}{k_B T}\right)$

As a result an average drift velocity $v_d$ arises in the direction of the electric force:

(3) $|v_d| = \lambda |r_{1\to 2} - r_{2\to 1}| = 2\lambda v_0 \exp(-U_a/k_B T) \sinh(2e\lambda E/k_B T) \cong$
$\lambda v_0 \exp\left(\frac{2e\lambda |U|/d - U_a}{k_B T}\right)$

Last approximation is valid when $2e\lambda E > k_B T$ what is completely fulfilled at studied intensities up to $10^7$ V/cm. At low voltage biases the mobility of ions is several orders of magnitude lower then effective mobility of charge carriers passing the interface. Therefore, under the threshold switching voltage when current density is typically below 10 ÷ 100 mA.mm$^{-2}$ (in ours as in Acha's experiments) we measure only the electronic conductivity. When the drift velocity of oxygen ions exceeds a critical value ($v_c$), instabilities in the ion system can take place similar to electrical instabilities in semiconductors. The observations are very similar to the known electrical instabilities in semiconductors driven far from thermodynamic equilibrium by strong external fields causing switching between high-resistive and low-resistive states[26]. If existence of a critical drift velocity $v_c$ is supposed to be a threshold for occurrence of the switching phenomenon then we can derive the threshold voltage $U_S$ defined as:

(4) $U_S = U|_{v_d \stackrel{\text{def}}{=} v_c} \cong \frac{d}{2e\lambda}\left\{U_a - k_B T \ln\frac{\lambda v_0}{v_c}\right\}$

Then, the threshold voltage at the zero temperature simply means the voltage level at which the kinetic energy of the ion gained in the electrical field equals the activation energy $U_a$ for movement from one unit cell to another. In that case, Eq. (4) simplifies to $\lambda e U_S/d \approx U_a$. Rising the absolute temperature, effective activation energy is reduced by the factor $k_B T \ln\frac{\lambda v_0}{v_c}$ and the threshold voltage is decreasing correspondingly. Such model gives linear relation between the switching voltage $U_S$ and the absolute temperature with the main parameters: the oxygen mean free path $\lambda$ and the activation energy $U_a$. Parameters $v_c$, $v_0$ occurs under logarithm therefore they are of second order of importance and their approximation in orders of magnitude is sufficient according to roughness of the model. Attempt frequency $v_0$ is typically in order of $10^{12}$ Hz[27]. Using the approximation for the mobile oxygen content in the barrier layer with the assumption of the one mobile oxygen per unit cell ($N \sim 5.10^{27}$ m$^{-3}$) we have estimated the critical drift velocity $v_c \sim 0{,}1 \div 1$ mm/s for entire temperature range from measured critical current densities $j_c$ and relation $j_c = Nev_c$.

Fitting the linear part of the temperature dependence over 200 K (Fig.5), we have received the model parameters: $\lambda = 4$ Å and $U_a = 0.7$ eV. Obtained value of the activation energy for the oxygen motion $U_a$ is in good agreement with previous works[15,24,28-30]. It is interesting to take note of the deviation from the linear law below 200 K. According to previous research this is a typical temperature below which the pseudo-gap and associated effects based on the YBCO antiferromagnetic nature can be observed [11,18,31]. In this state the spin correlation interactions are much stronger causing additional oxygen ordering in the CuO chains. The transition into a more ordered state increases the activation energy for the oxygen motion and reduces the oxygen diffusion. Therefore, also the electromigration based on the oxygen ions diffusivity is strongly suppressed what demands stronger electrical field for occurrence of the RS effect.

**Conclusion**

RS effect observed on metal/YBCO junctions in the temperature range 100-440 K was discussed within the electro-diffusion model, taking into account oxygen ions drift in the nanometer scale vicinity of the junction interface under applied electrical fields. The model leads to linear relation between switching voltage and the temperature while the measured dependence deviates from the linear law under 200 K. This point is worth of further investigation as the similar effect was observed also on our PtIr/YBCO contacts[14]. It seems that the interconnection between effects of the strongly correlated systems and the role of the oxygen electromigration is more striking below this temperature.